# A Review on Method Entities in the Academic Literature: Extraction, Evaluation, and Application

Yuzhuo Wang, Chengzhi Zhang [*]

Department of Information Management, Nanjing University of Science and Technology, Nanjing, 210094, China

Kai Li

School of Information Resource Management, Renmin University of China, Beijing, 100872, China

**Abstract:** In scientific research, the method is an indispensable means to solve scientific problems and a critical research object. With the advancement of sciences, many scientific methods are being proposed, modified, and used in academic literature. The authors describe details of the method in the abstract and body text, and key entities in academic literature reflecting names of the method are called method entities. Exploring diverse method entities in a tremendous amount of academic literature helps scholars understand existing methods, select the appropriate method for research tasks, and propose new methods. Furthermore, the evolution of method entities can reveal the development of a discipline and facilitate knowledge discovery. Therefore, this article offers a systematic review of methodological and empirical works focusing on extracting method entities from full-text academic literature and efforts to build knowledge services using these extracted method entities. Definitions of key concepts involved in this review were first proposed. Based on these definitions, we systematically reviewed the approaches and indicators to extract and evaluate method entities, with a strong focus on the pros and cons of each approach. We also surveyed how extracted method entities are used to build new applications. Finally, limitations in existing works as well as potential next steps were discussed.

**Key words: Method entities, entity extraction, entity evaluation, entity platform.**

## 1. Introduction

As an increasing amount of data is gathered to support scientific discoveries, a new scientific paradigm, or the Fourth Paradigm, emerged in the 21st century (Bell, Hey & Szalay, 2009). Under this new paradigm, scientific research becomes more data-driven, but there is still controversy over what is data-driven science. No matter the data-driven is regarded as a revolution in the history of science that overturns traditional science built around testable hypotheses ( Anderson,2008), or the scholar believes that natural historical sciences have fundamentally been data-driven sciences which have always rested on a combination of hypothesis-driven and data-driven methods (Strasser, 2012). There is no doubt that scientists face massive amounts of data in today’s data-driven science, and the growing amount of data highlights the importance of methods to scientific discoveries (Miller & Goodchild, 2015). While statistical methods for data analysis is the most frequently discussed category in the literature, researchers are also increasingly reliant on newer methods to support the entire data lifecycle, ranging from the capture, processing, curation, and visualization of data (Hey, Tansley & Tolle, 2009; Kelling et al., 2009).

[*] Corresponding author, Email: zhangcz@njust.edu.cn

Methods enable and help scientists carry out their research (Eales, Pinney, Stevens & Robertson, 2008), but finding or proposing a new method to adapt to the new scientific paradigm is not an easy task, which requires scholars to collect as many existing methods as possible and get a comprehensive understanding. In a given field, a sufficient number of methods enables scholars to analyze the characteristics of existing methods, and the newer method could be found or proposed based on the old methods. For the newer method, comparing its performance or contributions with other existing methods determines whether the new method will be recognized by the field. Based on all the old and new methods, experts can establish a method landscape for a discipline (Buitelaar & Eigner, 2009), which is beneficial to researchers, especially newcomers. In addition to maintaining an informed view of methods, the method evolution can be used to analyze the progress of a discipline. According to the method trend in a discipline, researchers can infer the changes in the research tasks and then get the development of research topics (Zhang, Tam & Cox, 2021).

For scholars who want to seek interdisciplinary cooperation, the method landscape of other disciplines may inspire their corporations. The more methods they get, the more attempts to solve problems with other disciplines' methods. Furthermore, if the mode of interdisciplinary cooperation based on methods was discovered, we can also recommend potential collaborators from other disciplines to scholars. Similarly, other knowledge services also require comprehensive collection and evaluation of methods. The academic search engine with the method search function needs to identify the methods in many academic papers and evaluate the matching degree between each method and the user's query. Scientific evaluation can also be carried out based on method use information, which can be used as a new indicator to evaluate the scientific influence of academic publications and researchers.

In summary, comprehensive collection, evaluation, and application of methods in a given discipline are essential for knowledge discovery, discipline development, and scientific cooperation. In scientific research, academic literature is an important resource for obtaining methods, and analyzing as much literature as possible can ensure the richness of the obtained methods. However, in this data-intensive era, a tremendous amount of literature is published every day (Jinha, 2010; Hassan et al., 2018; Xie et al., 2020). The traditional approach of manually exploring the related publications to find methods is becoming challenging because most scholars have never been able to keep completely up-to-date with publications considering the unending increase in quantity and diversity of research within their areas of focus (Ding& Stirling, 2016). Therefore, it is indispensable to conduct the study of extracting and evaluating methods from large-scale academic literature. It saves researchers labor and time, helping scholars form a method system, carry out research work, explore the evolution of disciplines, and expand knowledge services.

Therefore, this article plans to review the extraction and evaluation of methods in academic literature and the application of these extracted methods. Based on the current work, we pay more attention to the research on the names of the method, namely, method entities in academic literature (see the specific definition of the method entity in section 2), since studies on the entity are more easily extended to follow-up evaluation and applications. Generally, we aim at exemplifying the value of research on method entities in academic literature. We summarize the approaches for method entity extraction and indicators of method entity evaluation, with a discussion on each approach's strengths and weaknesses. We also review applications built on the method entities and related information. Additionally, limitations and potential future work are provided.

# 2. Definition of key concepts

This section defines key concepts involved in this literature review, including method entity, method entity extraction, and method entity evaluation.

## 2.1 Definition of the method entity in academic literature

In this literature review, method entities are regarded as the representation of *scientific methods* (or a part of the methods) as *named entities*. We will thus discuss how the concepts of scientific methods and name entities are approached in this literature review.

**(1) Definitions of methods and scientific methods**

As a commonly used concept, the definitions of methods or scientific methods are highly variables and subject to various contexts in which these concepts are used, such as time, disciplinarity, and geographical location.

*Merriam-Webster Dictionary* (2002) defines the method as a procedure or process for attaining an object. *Stanford Encyclopedia of Philosophy* illustrates that methods or scientific methods are the means of how the aims and products of science are achieved, which should be distinguished from meta-methodology and the detailed and contextual practices through which methods are implemented (Hepburn & Andersen, 2021). In the history of science, many scholars provided their insights on methods or scientific methods. Aristotle is recognized as giving the earliest systematic treatise on the nature of the scientific inquiry. In Aristotle's view, the methods were determined by the aims of discovery, ordering, display of facts, and include careful observation and logic as a system of reasoning (Gauch & Hugh, 2003). During the 16th-18th centuries, mathematical description and mechanical explanation were incorporated into the scientific method. Bacon's new method was processing results of systematic experiments and observations to discover the "form natures" of things (Bacon, 1878). Newton's method had two main thrusts: the implicit method of the experiments and reasoning presented in the *Opticks* (Newton, 1730) and the explicit methodological rules exemplified by the *Principia* (Newton, 1802). In the 19$^{th}$ century, the debate about the method between John Stuart Mill and William Whewell was characterized as a debate between inductivism and hypothetical deductivism (Macleod, 2020). After that, scientists in the 20$^{th}$ century believed that the crucial aspects of the scientific method were the means of testing and confirming theories. In recent years, the latest form of method studies have turned to "practice," which means methods are the detailed and context-specific problem-solving procedure and methodological analyses to be at the same time descriptive, critical, and advisory (Pitt & Pera, 2012). In general, it is a big challenge to give a clear definition of the method. Scholars in different eras have their own various opinions on methods. When they try to define methods, the definition depends on the goals for which the method is used or proposed. If the goal is to discover knowledge, the method is the observation, reasoning, and experiment in the discovery process. If the purpose is to solve the problem, then the method is testing, confirmation, and practice for problem-solving. At the same time, the requirements of different times and the development of science also affect the definition of methods. The earliest methods focused on observation and reasoning, then evolved into hypothetical deduction, and now are the means or ways to handle large amounts of data because the paradigm of scientific research has gradually evolved from hypothesis-driven to data-driven.

Since the definition of the method are subject to the context where methods appear, it can be speculated that the methods in the academic literature are different from general methods or scientific methods. Some scholars believe that academic literature does not reflect the complete process of obtaining scientific results, and it is a partial representation of what actually happened in the laboratory (Medawar, 1990). In the same way, the methods in academic literature do not show the formal logic of scientific methods, but rather an "opportunistic logic" of artful, situated practices (Woolgar, 1983). In the paper, a method has no dynamic structure of its own: no problems, no resources to transform problems, no fusion or fission of interests to carry out the operations (Knorr-Cetina, 2013). The authors may only choose the most essential and relevant objects or the describable content about a method involved in the research, for example, the name of a method, the brand of instruments, lists of materials, and descriptions of procedures tied together by sequence.

According to the introduction above, we recognize that how "methods" in the academic literature are defined is a uniform work. But it is deemed that methods in academic literature include the following objects: (1) the embodiment of some methods, or a part of the methods, or a package of several methods, and (2) frequently mentioned in the literature as what was used in the scientific research. In this literature review, we also aim at the methods obtained from academic literature. Taking into account the trend of method definition and the characteristics of the method in academic literature, we will define the method from the perspective of problem-solving practice. At the same time, we also consider the method objects that are often mentioned in the current scientific research to solve the problems proposed by authors, including discipline-specific methods, software, models, algorithms, and metrics. Generally, in this review, the methods in the academic literature are *the ways, means, and channels used to solve tasks or problems proposed by the authors, including sub-categories such as discipline-specific methods, software, models, algorithms, and metrics.*

**(2) Definition of method entities**

In academic literature, methods are mentioned in various forms. They can be sentences about the "methods discourse" representing the methodological statements, explanations of methodological concepts, methodological imperatives, and justifications of experimentation strategies. They can also be the terminology terms indicating the name of specific methods (Schickore, 2017). Compared with method discourse, the method term in the content of academic literature has more explicit boundaries, and it can be regarded as a kind of named entity.

The named entity (NE) was first proposed in the Sixth Message Understanding Conference (MUC-6) in 1995, but the definition of the named entity was not discussed in the conference. The organizers only explained that it was a "unique identifier of the entity" that could be divided into names of persons, organizations, and places (Appelt et al., 1995). After that, CoNLL-2002 (Tjong Kim Sang, 2002) and CoNLL-2003 (Tjong Kim Sang & De Meulder, 2003) defined named entities as phrases containing names, including person name, place name, organization name, time, and quantity. In addition to conferences, scholars also discussed the meaning and types of named entities. Petasis et al. (2000) believed that named entities were proper nouns (PN), which served as the name of someone or something.

From the perspective of ontology, Alfonseca and Manandhar (2002) proposed that named entities were objects used to solve specific problems. Borrega et al. ( 2007) defined named entities in detail from the perspective of linguistics, stipulating that only nouns and noun phrases can be used as named entities. Although these definitions are not uniform, it can be sure that named entities are nouns or noun phrases indicating unique things.

Based on the above definitions, we define method entities as named entities that represent specific methods. In particular, method entities in the academic literature are *nouns or noun phrases representing the specific ways, means, and channels used to solve tasks or problems proposed by the authors, including sub-categories such as discipline-specific methods, software, models, algorithms, and metrics.*

## 2.2 Definition of the method entity extraction and evaluation in academic literature

Method entity extraction can be regarded as a sub-task of named entity recognition. Named-entity recognition (NER) (also known as (named) entity identification, entity chunking, and entity extraction) is a subtask of information extraction that points to identify and classify members of rigid designators from data suited to different types of named entities such as organizations, persons, locations, et al. (Goyal, Gupta & Kumar 2018). Method entity extraction means that the object of NER becomes the method. It needs to identify nouns or noun phrases representing a method from unstructured text. It can also be considered that this task is to find the name of methods from the text. Furthermore, method entity recognition can also be extended to method entity recognition and classification; method entities are further divided into different method objects. Generally, method entity recognition can use expert experience, the linguistic knowledge related to the method entity, the characteristics of the method entity and even the context information to identify and classify the noun or nun phrase representing a method from the text.

The scope of the method entity evaluation can be the popularity, influence, complexity, quality, and performance of the method in specific tasks or fields. For method entities obtained from large-scale literature, the evaluation focus on the influence of method entities. Ding et al. (2013) proposed entity metrics to measure the influence of entities, which aims to demonstrate how bibliometric approaches can be applied to knowledge entities and ultimately contribute to knowledge discovery. In Ding's framework of knowledge entities, method entities belong to micro-level entities. Therefore, the evaluation of method entities' influence can also be conducted by bibliometric approaches and altmetric approaches, which means scholars measure method entities by different kinds of frequency, and then get the rank of methods in a specific field or task. Furthermore, combining the frequency with other features, method entities can be analyzed from the temporal perspective to capture dynamic changes or from the spatial dimension to identify geographical differences. Method entity evaluation helps scholars better understand the difference in influence between method entities, and get insights into how different characteristics and environments affect the influence of method entities.

# 3. Collection of related literature

Following the above definition, we listed some nouns related to the research objects of this review, including "*method entity, method knowledge element, method term, method name, technology, software, algorithm, model, data set and so on*", some words about the research field, such as *scientific research, academic literature, academic papers et al*., and words about the research topic, including *extract, assess, evaluate, apply and corresponding nouns*. The three types of words were combined as search terms and searched on the *Web of Science, Google Scholar, Springer Link, and ScienceDirect*. By browsing the title and abstract of each article, we screened out relevant documents from the search results, and searched for related literature in the reference list when reviewing obtained articles. Finally, we obtained a total of 69 related publications about extracting method entities from academic literature, evaluating the influence of the method entities obtained from academic literature, and the application platforms of these method entities.

We reviewed the publication time of these 69 papers and counted the number of papers published each year. The related work first appeared in 1977, and we found only a few papers on this topic published in the 1990s, possibly due to the fact that there were not enough large-scale datasets to support this line of research. After 2010, the number of related papers published each year gradually increased, and the research objects became more method objects with detailed categories, like algorithms. We speculate it is because, with the access of full-text databases and the advancement of machine learning methods, research on method entities acquired more external resources. The number of publications did not rise significantly until 2016, when deep-learning technology has been widely used in method entity extraction, and the evaluation of method entity is becoming more diversified.

# 4. Identifying and extracting method entities from academic literature

Related work about method entity extraction is discussed in this section. Based on the process of method entity extraction, we introduce these related research from two aspects: corpus collection and entity extraction. Therefore, first, we briefly describe how the existing works obtained or constructed the data corpus for the method entity recognition tasks, and introduce some commonly used data sources. Next, we present various approaches to method entity extraction, including manual annotation, rule-based methods, statistical machine learning methods, and deep learning methods.

## 4.1 Collecting the corpus for method entity extraction

Datasets serve as the primary ingredient in method entity extraction, and they are the resources of various method entities. Commonly used data sets are divided into two formats: PDF and XML. For the data in PDF format, scholars can directly obtain method entities by using manual annotation. Nevertheless, the PDF data set is not machine-friendly; it needs to be converted into XML format or TXT format for automatic method entities extraction. In this case, optical character recognition (OCR) has become an indispensable technology for parsing texts from PDF files, but OCR would add noises to the extracted texts (Tuarob, 2014). We need more professional institutions to provide textual academic literature data. Some existing data sets offer academic literature in TXT and XML format, including bibliographic and full-text content. Although there are still some errors in these data, such as garbled characters and incomplete content, they provide strong support for method entity extraction.

The appendix shows the popular databases that provide data set for existing method entity extraction. Most databases provide full-text content of academic literature. ACL ARC, PLOS, and PubMed Central provide full-text data in XML format that machines can process directly. One point worth noting is that the documents provided by these databases mainly come from the domain of computer science, biology, and medicine.

## 4.2 Approaches to extracting method entities from academic literature

We identify the following four frequently utilized methods in method entity extraction: manual annotation, rule-based extraction, traditional machine learning, and deep learning. Related works using these approaches to extract method entities are discussed below.

### 4.2.1 Manual annotation of method entities

Manual annotation is the task of reading a particular preselected document and providing additional information in the form of the so-called annotations, which can occur at any level of a linguistic component, i.e., document, paragraph, sentence, phrase, word, or character (Neves & Ševa, 2021). As early as the last century, scholars began to identify methods in academic literature based on content analysis (Peritz, 1977; Blake, 1994; Jarvelin & Vakkari, 1990), but these works summarize the methods based on the contents of documents, paragraphs, and sentences, instead of getting the words or phrases that represent the method directly from the text. In contrast, manual annotation of method entities is the annotation in the phrase or word level. Organizers usually compile an annotation guide first and then recruit and train the labelers. After pre-labeling, the organizer decides whether to modify the annotation guideline. Finally, the annotator reads the academic literature content and annotates the phrase or word representing the method entity according to the annotation guide.

For method entities in academic literature, the source can be bibliographic content. QasemiZadeh and Schumann (2016) annotated the terms (i.e., single or multi-word lexical units with a specialized meaning) in abstract of academic papers, including disciplinary approaches, tools, models, measures, and measurements. However, due to space limitations, many authors do not describe their methods in the abstract; instead, they introduce their own proposed or used methods in the body text, especially the method section in academic papers, and mention some other methods as background knowledge. Compared, the full-text of academic literature contains more method entities and more information about the method entities, such as how these methods are used. Ibrahim (2021) analyzed the statistical methods entities used in academic papers in library and information science (LIS). In addition to the name of the statistical method, the section where the method is located, they also recorded whether the use of statistical methods was realized by software. Similarly, Howison and Bullard (2016) manually annotated the name of the software mentioned in the body text of papers. At the same time, they recorded the software's mention characteristics, for example, the URL, version number, creator and accessibility of software. Recording these attributes or features of method entities often need more human judgment, unlike the entity names, they cannot be directly copied from content. In this way, the advantage of manual annotation is reflected, that is, more complex information can be obtained by annotation.

However, it should be noted that the acquisition of complex information about entities puts forward higher requirements for the annotators. Specifically, manual coding requires the annotator to have valid knowledge of the labeled method to determine which part of the text is the method entity. However, each annotator's understanding of the method will lead to inconsistent annotation results. Therefore, when multiple labelers participate in the labeling task, they invite experienced experts to review the annotation results of ordinary students (Wang & Zhang, 2018a). In more cases, the measurement of interrater reliability between each annotator is needed to prove the validity and availability of labeling results. The common indexes are Cohen's kappa (McHugh, 2012) and Krippendorff's Alpha (Krippendorff, 2011). After measuring interrater reliability, the final labeling can be that a labeler completes all tasks (Wang & Zhang, 2020), or different labelers divide the work equally (Wang & Zhang, 2019), or different labelers annotate the same text and retain the method entities obtain the same label from each labeler (Hou et al., 2020). These measures guarantee the accuracy of the method entities; therefore, the method entities that are manually labeled become the indispensable training corpus for the automatic extraction models and the gold standard for judging the results of automatic extraction (Eales et al., 2008).

Whether it is to recognize method entities directly or prepare the training corpus for automatically extracting entities, manual labeling is an integral approach in method entity recognition. This approach ensures establishing an accurate set of baseline knowledge, including method entities and their essential attributes, through human judgment to support more large-scale analysis. However, manual annotation is a highly time-consuming and labor-intensive method that makes it challenging to process large-scale corpora. Therefore, most of the existing works using this approach only analyze hundreds of articles, which dramatically limits the generalizability of their findings.

### 4.2.2 Rule-based extraction methods

Rule-based method entity extraction is based on textual rules of how method entities are mentioned. Researchers need to identify these rules and then apply them to identify more entities and related information. Dictionary matching is a standard rule-based method, where researchers compile a dictionary of entity names through other resources and find such names in the text. More importantly, it also records additional valuable information about the entity in the context (Ding et al., 2013). For example, Wang and Zhang (2018b) constructed a dictionary of abbreviations and aliases of 10 popular text mining algorithm names from academic search engines and online sources. Using this dictionary, they found the ten algorithms in the NLP conference papers with an exact match, and the scientific tasks connected to these algorithms.

Dictionary-based methods can directly locate method entities mentioned in the text, but this approach has its limitations as well. The most significant limitation is that it can only acquire the entities and their name forms included in the dictionary, which excludes any other method entities and different name forms of existing entities. Previous studies have shown that the method entities identified by more complex rules meet searchers' needs better than those identified by matching search terms with academic papers (Bhatia, Mitra & Giles, 2010). Therefore, scholars tried to propose more complex rules, including the cue words, linguistic patterns, part of speech, location of words, *et al.* (Katsurai, 2021; Lam et al., 2016; Li & Yan, 2018; Wang & Zhang, 2019; Zhu et al., 2013). Combining dictionary matching and other rules, Duck et al. (2013) created bioNerDS, a bioinformatics named entity recognizer, to extract software entities and data set entities from papers. They obtained all noun phrases in academic papers and scored these phrases by different rules. In the first round, they generated a dictionary to check whether the candidate entities appeared in the dictionary. In the second round, strong rules about the local clues, like the version information, references, and URLs, were classified into positive and negative rules and assigned different scores. In the final round, some clues like specific verbs, indicative but ambiguous heads were combined as weak clues and assigned the score. Candidate entities were scored according to whether they met rules and judged as method entities by the final score.

Rule-based methods are considered highly efficient because they exploit the properties of language-related knowledge to obtain sufficient accuracy (Shaalan, 2010). However, making plenty of hand-crafted rules is expensive, which enables using iterative patterns to update rules dynamically and ensure the richness of rules. The iterative approach sets seed rules to find sentences containing method entities, from which newer rules and entities are extracted until no new knowledge can be acquired anymore (Zhang, Cheng & Lu, 2016). Gupta and Manning (2011) proved that the new rules obtained by the iterative approach achieved a higher F-1 value than the seed rules in technology entity (essentially the method entity in our definition) recognition.

Compared with manual annotation, the rule-based approach can support studies of larger scales, and it performs well on data from one specific field due to the manual effort and human observations. Simultaneously, the effectiveness of rule-based methods depends on the richness of rules, but making rules requires human expertise regarding knowledge of the domain and language and programming skills (Thenmalar, Balaji & Geetha, 2015), which is laborious and time expensive. At the same time, these expensive rules highly rely on domain knowledge and cannot be used in multiple domains conveniently. For example, if scholars want to use existing rules about English method entities to process Chinese academic literature, the efficiency of the rules will be significantly reduced. In general, rule-based approaches are found quite expensive, domain-specific, and non-portable. These limitations enable scholars to find more flexible means to extract method entities.

### 4.2.3 Statistical machine learning-based extraction methods

With the popularity of machine learning in natural language processing, this approach is also used to extract named entities from academic texts. Statistical machine learning approaches generally regard NER as a classification or sequence tagging task, some classical approaches, such as the hidden markov model (HMM) (Bikel, Miller, Schwartz, Weischedel, 1998; Bikel, Schwartz, Weischedel, 1999), conditional random field (CRF) (McCallumn & Li, 2003), maximum entropy (ME) (Borthwick & Grishman, 1999) and support vector machine (SVM) (Isozaki & Kazawa, 2002) have been successfully used for sequence tagging of entities, and gradually been used in method entity recognition. A single machine learning method, such as a perceptron, can achieve good results in method entity extraction (Tateisi et al., 2016). The success inspired scholars to continue to propose new optimization strategies to improve the performance of classic machine learning approaches and obtain better extraction results. Next, we will introduce several strategies for optimizing machine learning algorithms in method entity extraction.

The first common optimization strategy is to choose better features of method entities. Features are the properties and attributes of textual objects in a computational model, which are further used by the learning methods for generating a model (Goyal, Gupta & Kumar, 2018). A classical framework divides different features into list lookup features (additional information such as gazetteers or lexicons), document and corpus features (document structure and content), and word-based features (casing, part-of-speech tag) (Nadeau & Sekine, 2007). These features can be used together to improve one extraction model. For example, Wei et al. (2020) used the bag-of-word, word shape, morphological information, POS tag, domain knowledge, and word embedding features to optimize the performance of the CRF model in software entity extraction. Besides optimizing the single model, combining different features and models is another method used to acquire better extraction models (Liakata et al., 2012; Liakata & Soldatova, 2008). Researchers have used the following patterns. The first is to compare the models first, where the best model as well as features to fit the models are selected (Meng, Hou, Yang & Li, 2018). The other method is to directly combine different models and features to develop the best model-feature pair and use the model to extract the method entity (Heffernan & Teufel, 2018)

Another optimization strategy is integrating different extraction models. This strategy usually constructs a two-layer model to extract method entities. The first layer identifies method sentences in the full-text, and then the second layer extracts method entities from method sentences. The recognition of method sentences in first-level can be achieved by the rule-based method (Houngbo & Mercer, 2012) or multiple classification models, such as a support vector machine, k-nearest neighbor, decision tree, and naive Bayes (Kovačević et al.,2012). The entity extraction in the second-level is usually done by using the sequence labeling model, such as CRF, that considers multiple features. This hybrid method breaks down the method entity recognition into different subtasks so that the more suitable approach can be used in each step. At the same time, the practice of first identifying sentences and then identifying entities reduces the scope of entity identification, thereby improving the performance of statistical machine learning models.

The third optimization strategy is reducing the investment in the extraction process. To be more specific, this strategy uses semi-supervised machine learning approaches to reduce the need for expensive training data produced by human annotators in supervised machine learning models (Siddiqui, Ren, Parameswaran & Han, 2016). A few seed training examples are used to re-train the model to generate more labeled examples, continuing several rounds until no more new entities can be obtained (Pan, Yan, Wang & Hua, 2015). On the basis of semi-supervised, Mesbah, Lofi, Torre, Bozzon, and Houben (2018) presented TSE-NER, a CRF-based semi-supervised learning approach for extracting method entities from scientific publications with minimal human input. Several seed words and sentences were selected, and word2vec and doc2vec were used to develop training terms and sentences separately based on semantic similarity. Five strategies were presented to filter the candidate method entities. After that, Färber, Albers and Schüber (2020) modified the term expending strategy of TSE-NER by using SciBERT as a semantic relatedness method and k-means to cluster the new entities using. Generally, these works prove that the semi-supervised strategy minimize the training costs in scenarios where the extracted entity types are rare, which makes it possible to obtain a high-precision method entity extraction model with only a small amount of seed words.

On the premise that large-scale data can be processed, machine learning provides a more flexible and widely applicable way for method entity extraction. Recent developments have greatly improved the accuracy of extraction. However, machine learning-based methods rely on annually labeled training corpora and feature engineering. It still takes a lot of time and labor to construct a high-quality labeled corpus and features, which remains to be further improved by future works.

### 4.2.4 Method Entity extraction based on deep learning methods

With the development of technology, the deep learning methods derived from the neural network model are becoming a hot topic in machine learning. Compared with the statistical machine learning models, deep learning models no longer rely on complex features, significantly saving the workforce and time spent on feature engineering. Therefore, Deep learning methods are also recognized by more scholars and used to extract method entities from academic literature.

Works have been done to compare the deep learning model with statistical machine learning models in method entity extraction (Jiang, Zhu, Shen, Wang & Catalano, 2019). The comparison between SVM, CNN, LSTM, RCNN, Bi-LSTM indicates that the character vector-based Bi-LSTM model achieved better performance in identifying a series of scientific entities, including many method entities covered by our definition (Zhao et al., 2019). More work is to combine Bi-LSTM with other models. Although Bi-LSTM can learn the context information of the entity, it does not consider the relationship between the output results. Therefore, the CRF model is used with Bi-LSTM. By calculating the relationship between adjacent output tags, CRF could limit and filter the results provided by Bi-LSTM to obtain more accurate extraction results. (Lei and Wang, 2019). Moreover, some scholars found that using the convolutional neural network (CNN) to capture the character-level features of the training corpus can improve the performance of sequence annotation and other tasks. Therefore, the CNN model is combined with CRF + bi LSTM to extract method entities (Ammar, Peters, Bhagavatula & Power, 2017; Chen, Trabelsi, Davison & Heflin, 2020). Yao et al. (2020) built the CNN+ Bi-LSTM+CRF model to extract method entities from AI academic papers. The character-level and word-level embeddings were obtained through the CNN network and Glove, and then two kinds of embeddings were co-created and fed into Bi-LSTM with association amongst words calculated by self-attention. Finally, the hidden vector was sent to the CRF and obtained the method entities from sentences. The improved model achieved good results where the F-1 value increased to 0.792.

In addition to the deep learning models introduced above, a newer deep learning model, BERT (Devlin et al., 2019), is also widely used in method entity extraction because of its superior performance. Jiang et al. (2019) used BERT to extract software entities from full-text papers and achieved an F-1 value of 85.44%. Zheng et al. (2021) chose another special BERT trained on scientific text, SciBERT, to extract general method entities and achieved an F-1 value of 81.4%. Although we cannot directly compare the performance of BERT and SCIBERT with these two works because the software entity and the general method entity are at different granularities, it is evident that various BERT performs better than the previously introduced deep learning model in extracting all kinds of method entities.

Although deep learning methods no longer rely on feature engineering, they still require large-scale annotated training corpora. However, there are few public data sets available for method entity extraction. To obtain enough data for deep learning model training, weakly-supervised deep learning models, including semi-supervised (Luan, 2018) and distant supervised (Boland and Krüger, 2019), are applied to reduce the dependence on the annotated corpus. For the semi-supervised algorithm, Luan et al. (2017; 2018) introduced a graph-based algorithm to estimate the posterior probabilities of unlabeled data and used CRF to take the uncertainty of labels into account while optimizing the objective function. They found that the LSTM-CRF in semi-supervised learning significantly outperforms the supervised tagging model in the task of method entity extraction. For the distant supervision, Jain, Van, Hajishirzi, and Beltagy (2020) expended the training dataset based on entities provided by Paper with Code corpus[1]and then used BERT-BiLSTM-CRF to extract multiple entities, including method. Weak supervision and distant supervision can also be combined to solve the same task. For example, Schindler et al. (2020) produced a small gold standard corpus (GSC) by manual annotation and then built a large-scale silver standard corpus (SSC) based on the weak supervision tool (Snorkel data programming framework) and distant supervision (Wikipedia). The Bi-LSTM-CRF model trained by the SSC+GSC achieved better performance in software name recognition than that of the model trained by SSC or GSC individually.

[1] https://www.paperswithcode.com/

On the whole, the deep learning model is becoming more of scholars' choice to extract method entities. Compared with a single model, the combination of statistical machine learning model and deep learning model has become more popular because developing new technologies using the most vital points of each one can improve the model's effectiveness. However, when using the deep learning model to extract method entities, we still need to pay attention to reducing the dependence of the model on large-scale training corpus or reducing the cost of obtaining large-scale training corpus while ensuring the overall performance of the model. Methods such as the CRF model (Jie, Xie, Lu, Ding & Li, 2019), EM algorithm (Paul, Singh, Hedderich & Klakow, 2019), and reinforcement learning (Yang, Chen, Li, He & Zhang, 2018) have been used to improve the quality of training data. In the future, it is bound to be used to optimize the extraction model of method entities.

### 4.2.5 The summary of approaches to method entity extraction

Table 1 summarizes method entity extraction literature covered in this review. In general, the approach of method entity extraction has experienced the evolution from manual annotation, rule-based method, and statistical machine learning to deep learning models, which follows the development of the NER approach.

According to table 1, it is difficult to say which approach is the best for method entity identification because the method objects extracted in each work are not consistent, and the data source and data scale are not identical, which would affect the performance of different methods in the task. In any case, selecting an appropriate approach is a big challenge for method entity extraction since each approach has advantages and disadvantages. To be specific, the accuracy of manual labeling is high, but the work efficiency is low. The rule-based method can deal with more data, but it has low flexibility and can only be used in specific fields. Statistical machine learning is more flexible in dealing with larger-scale data but relies on expensive feature engineering. The deep learning method seems to solve all the shortcomings of the above methods, but it still faces the dependence on large-scale training corpus in essence and the need to further improve the accuracy compared with manual annotation. From the development trend of existing methods, it seems that the hybrid approach, which integrates different extraction methods, will become the choice of more people. The latest hybrid approaches use manual annotation to build a small-scale gold-standard corpus, and use the rule-based method combined with external resources to generate more training corpora. These corpora are then utilized for training the extraction model combined with statistical machine learning and deep learning models. This pattern obtains the advantages of all kinds of extraction methods, but some aspects still need to be improved. For example, how to use small samples for training, control the quality of non-manually annotated corpus in weakly supervised learning, make the model can be used for different method objects, and make the model available in different languages and disciplines. In other words, we hope to get a simple, accurate, flexible, low-cost approach that can be used in different fields in the future.

**Table 1. Method entity extraction summary.** **(1)**The papers corresponding to each ID in the appendix. QasemiZadeh and Schumann (2016); 2. Ibrahim (2021); 3. Howison and Bullard (2016); 4. Wang and Zhang (2018a);

5. Wang and Zhang (2019); 6. Wang and Zhang (2018); 7. Katsurai (2021); 8.Katsurai (2021); 9. Li and Yan (2018); 10.Duck et al. (2013) 11. Gupta and Manning (2011); 12. Lam et al., (2016); 13. Tateisi et al., (2016); 14. Wei et al., (2020); 15. Meng et al., (2018); 16. Heffernan and Teufel (2018); 17. Houngbo and Mercer (2012); 18. Kovačević et al. (2012); 19. Mesbah et al., (2018); 20. Michael et al., (2020); 21.Jiang et al., (2019); 22.Zhao et al., (2019); 23. Lei and Wang (2019); 24.Ammar et al. (2017); 25.Yao et al., (2020); 26. Zheng et al., (2021); 27. Luan et al.(2017, 2018); 28.Jain et al. (2020); 29. Schindler et al. (2020). **(2)** The work whose "type" is shown as "method" in the table means that the author only extracts method entities without further categorizing method entities. If the method entity and other entities are extracted at the same time in the related work, as long as the individual extraction result of the method entity is provided, only the result of the method entity will appear in the table. **(3)** “***” means the best model or feature.

| ID | Extraction methods | Rules/Features/ Properties | Types of entity | Data set | Results | | |
|---|---|---|---|---|---|---|---|
| | | | | | Agreement (%) | | |
| 1 | Manual annotation | —— | Terms, including the method entities | Abstracts of 300 papers from the ACL Anthology Reference Corpus (ACL-ARC) | Cohen's kappa：Iteration 1: 76.00；iteration 2: 78.70；iteration 3: 60.80；iteration 4: 67.00；F-1 value: 72.70 | | |
| 2 | Manual annotation | —— | Statistical methods | Full-text content of 881 papers in LIS domain | —— | | |
| 3 | Manual annotation | —— | software | Full-text content of 90 papers in the Biology domain from WOS | Byrt's kappa: 100.00 | | |
| 4 | Manual annotation | —— | Method | Full-text content of 2,450 papers from PLOS One | —— | | |
| 5 | Manual annotation | —— | Algorithms | Full-text content of 4,641 papers from ACL Anthology | Cohen's kappa:78.00 | | |
| | | | | | **Precision (%)** | **Recall (%)** | **F-value (%)** |
| 6 | rule-based method | Dictionary | Algorithms | Full-text content of 5,212 papers from ACL Anthology | —— | —— | —— |
| 7 | rule-based method | Cue word | Method | Full-text content of 4,568 papers from ACL Anthology | —— | —— | —— |
| 8 | rule-based method | N-gram, frequency of word, cue word | Data Mining Methods | Title of 16,029 research papers in the area of data mining and analysis from Scopus | —— | —— | —— |
| 9 | rule-based method | linguistic patterns | software | Full-text content of 13,684 papers from PLOS | 95.00 | 93.50 | —— |
| 10 | rule-based method | Dictionary, local clues,Cross-mention "weak" clues | software | Full-text content of 60 papers from BMC Bioinformatics and PLoS Computational Biology and Genome Biology corpus | 82.00 | 73.00 | 77.00 |
| 11 | rule-based method | Semantic patterns | Technology, including the method entity | Titles and abstracts of 15,016 papers from the ACL Anthology Network and ACL-ARC | 30.46 | 46.68 | 36.86 |
| 12 | rule-based method | Part-of-speech, lexicon，syntax | Methodology terms | Method and results sections of 3,720 papers from the domain of sleep disorders | 70.00 | 77.00 | 75.51 |
| 13 | Perceptron | —— | Processual Entity (method) | Abstracts of 250 papers from the ACL anthology and 150 papers from the ACM digital library | 64.00 | 66.00 | 65.00 |
| 14 | CRF | **Basic features：** (1)word shape (2)n-gram (3)sentence (4)prefix–suffix **Domain knowledge features：** (1)section (2)dictionary (3)orthographic feature **Unsupervised word representation features:** (1)discrete word embedding (2)clustering of word embedding*** | Software | Titles and abstracts of 1,120 papers from the PubMed | 88.89 | 76.92 | 82.47 |
| 15 | NB | —— | Metadata, including the method entity | Full-text content of 30 papers from ACL Anthology | 51.80 | —— | —— |
| | K-NN | —— | | | 63.36 | —— | —— |
| | Adaboost | —— | | | 62.28 | —— | |
| | DT | —— | | | 54.61 | —— | —— |
| | LR | —— | | | 57.54 | —— | —— |
| | RF *** | **Basic features:** (1)frequency (2)max word length (4)leading letter capitalized (5) location (6)lexical cohesion | | | 64.44 | —— | —— |
| | | **Fine-grained distribution features:** Importance of section | | | 58.58 | —— | —— |
| | | All features | | | 67.63 | —— | —— |
| 16 | NB | (1)bags of words (2)Transitivity (3)Modality (4)Polarity (5)Syntax(6)Doc2Vec (7)Word2Vec (8)$Word2Vec_{smoothed}$ | Method | Full-text content of 2,500 papers from ACL Anthology | 75.5 | —— | —— |
| | SVM | (1)bags of words (2)Transitivity (3)Modality (4)Polarity (5)Syntax (6)Doc2Vec (7)Word2Vec (8)$Word2Vec_{smoothed}$ | | | 80.1 | —— | —— |
| | LR | (1)bags of words (2)Transitivity (3)Modality (4)Polarity (5)Syntax (6)Doc2Vec (7)Word2Vec | | | 81.5 | —— | —— |
| 17 | Sentences classification: Rule-based method | —— | Method | 6500 pairs of sentences from 189 different journals and 2000 papers in PubMed | 85.4 | 100 | 91.89 |
| | Entity extraction: CRF | (1)word feature (2)part-of-speech tags (3)word-shape features (4)position features (5)token prefixes and suffixes features (6)bigram features | | | 81.8 | 75 | 78.26 |
| 18 | Sentences classification: SVM, K-NN, DT, NB*** | —— | —— | —— | 74.00 | 82.00 | 78.00 |
| | Entity extraction: CRF | **Lexical features:** (1)phrase (2)normalize phrase (3)phrase type **Frequency features:** (1)Frequency of entity **Syntactic features**: (1)dependent relation (2) governor relation **Others:** Citation feature, category of the verb | Method, task, resource, implementation | Full-text content of 110 papers from Association for Computational Linguistic (ACL),Association for Computing Machinery and Conference on Computational Linguistics | 70.46 | 42.51 | 53.03 |
| 19 | CRF | **Expansion strategy:**(1)word2vec(2)Doc2vec*** **Filtering strategy:**(1)WordNet + stopwords (2)similar terms (3) pointwise mutual information (4)knowledge base lookup (5)ensemble **No filtering*** ** | Method, dataset | Full-text content of 15,994 papers from eight conferences in the domain of data science and processing | 68.00 | 15.00 | 25.00 |
| 20 | CRF | **Expansion strategy:** SciBERT, k-means **Filtering strategy:**(1)parts-of-speech analysis (2) stop-word removal (3)knowledge graph information (4)similarity scores | Method, dataset | SciREX (Jain et al., 2020) | 44.00 | 14.00 | 21.00 |
| 21 | CRF++ | —— | software | Full-text content of 892 papers from Journal of the Association for Information Science and technology | 95.12 | 78.68 | 86.12 |
| | BiGRU-CRF | | | | 89.50 | 87.37 | 88.40 |
| | BERT | | | | 85.81 | 85.10 | 85.44 |
| 22 | AvrgEmbed+LR | —— | scientific entities, including method entities | Full-text content of 52,705 papers from ACL Anthology Reference Corpus, NIPS Proceedings, and the PubMed | —— | —— | 45.30 |
| | AvrgEmbed+SVM | | | | | | 42.90 |
| | CNN | | | | | | 45.00 |
| | LSTM | | | | | | 41.30 |
| | RCNN | | | | | | 43.00 |
| | FastText | | | | | | 48.90 |
| | character vector-based Bi-LSTM | | | | | | 53.20 |
| 23 | CRF | —— | Models | Full-text content of 893 papers from Journal of the Association for Information Science and technology | 91.32 | 69.02 | 78.62 |
| | Bi-LSTM+CRF | | | | 81.16 | 79.39 | 80.26 |
| 24 | CNN+Bi-LSTM+CRF | —— | Process (method), task, material | SemEval-2017 Task 10 SCIENCEIE(Augenstein et al., 2017) | | | 55.20 |
| 25 | CNN+BiLSTM+CRF | —— | Method | Full-text content of 122,446 papers from list of AI journals and conferences in China Computer Federation (CCF) | 77.30 | 76.10 | 78.00 |
| | rule-based method +CNN+BiLSTM+CRF | | | | 79.10 | 80.40 | 79.10 |
| 26 | SciBERT | —— | Method | 2000 sentences in papers from ACL ARC annotated by the authors | —— | —— | 81.40 |
| 27 | CRF+LSTM | **External resources for pretraining: :**(1) Wikipedia (2) ACM | Process (method), task, material | SemEval Task 10 SCIENCEIE | —— | —— | 42.90 |
| | | **Semi-supervised Learning:** CRF training with uncertain labels<br>**External resources:**(1) Wikipedia (2) ACM | | | —— | —— | 44.40 |
| | | **Semi-supervised Learning:** Graph-based Posterior Estimates<br>**External resources:**(1) Wikipedia (2) ACM | | | —— | —— | 43.3 |
| | | **Semi-supervised Learning:** Graph-based Posterior Estimates+CRF training with Uncertain Labels<br>**External resources:**(1) Wikipedia (2) ACM | | | —— | —— | 45.3 |
| 28 | DYGIE++ | External resources: Paper with code<br>Noisy labeling: BERT+CRF trained on SCIERC dataset<br>Human corrections | Dataset, metric, task, method | Using SCIERC data set to generate SCIREX | 63.70 | 64.00 | 63.8.00 |
| | BERT+Bi-LSTM+BIOUL based CRF | | | | 67.60 | 69.40 | 68.50 |
| 29 | Bi-LSTM+CRF | **gold standard corpus (GSC):**<br>manual annotation | Software | Content of 51,165 Methods & Materials (M&M) sections of papers from PloS | 20.00 | 68.00 | 30.00 |
| | | **silver standard corpus (SSC):**<br>External resources: Wikipedia<br>Noisy labeling: Snorkel data programming framework | | | 80.00 | 72.00 | 76.00 |
| | | **GSC+SSC** | | | 83.00 | 82.00 | 83.00 |

# 5. Evaluating the influence of method entities in academic literature

After obtaining method entities from large-scale academic literature in different ways, evaluation is needed to measure the entities' role, value, and influence to help scholars better understand the obtained method entities. Traditional evaluation of method entities pays attention to the quality evaluation based on scoring frameworks (Boehm, 1991; McCall, Richards & Walters, 1977; ISO/IEC 9126-1, 2001; ISO/IEC 25010, 2017) and conducting specific experiments (Eckle-Kohler et al., 2013; Wilbanks, Facciotti & Veenstra, 2010; Zhang, Jiang & Li, 2016). In this review, we focus on the impact evaluation of method entities because influence is an expression of recognition that can indicate the quality and value of a method. At the same time, impact evaluation is a classic research frame in quantitative science studies, and it illustrates how the method entities are used in scientific research and practices. This section introduces the related work about the influence assessment of method entities. We divided these researches into two parts, the first is the academic influence of method entities, and the second is the social influence of method entities.

## 5.1 Academic influence evaluation of method entities

The academic influence of method entities is usually evaluated by the bibliometric indicators, including the number of citations, mentions, and usage of entities in academic literature. Among the various indicators, citation counts have been used for scientific assessment with a long history (Garfield, 1955). Citation number is the basis of SCI and contemporary evaluative scientometrics; it reflects the recognition from qualified peers and can be seemed as a non-financial reward (Garfield, 1979). Under this paradigm that is still having strong impacts on contemporary quantitative science studies, citation count is an inevitable instrument that researchers use to measure the influence of method entities. Based on the number of citation sentences, Ding, Wang, and Zhang (2019) evaluated the influence of algorithms used in the community of natural language processing. However, citation count has been proven to be an imperfect measurement for the impact of method entities in the following scenarios. The first is that the article mentioned and used the method entity but does not give any labels for citing the entity. If we only use the number of sentences or articles citing the method entity to measure the impact, there will surely be an inaccurate result. The second is that the literature cited the method entity with a non-standard reference, for example, citing the method entity by footnotes or in-text URLs (Howison, Deelman, McLennan, Silva & Herbsleb, 2015). If only standard citations are considered when counting citations to entities, the entity's influence will be weakened. The third case is that the evaluated method entity contains multiple different cited objects. For example, Li, Chen, and Yan (2019) pointed out that the citable objects of the software include not only software papers and software project pages, but also unpublished manuscripts created throughout the history of the software package. If we only count one of these citable objects, the impact of software will be significantly underestimated.

Therefore, the number of mentions and the amount of use are selected as implement indicators, which can be used to measure method entities that have not been formally cited but contributed to the literature. Similar to the citation count, the units of mention count and usage count are divided into the number of documents and the number of sentences. Wang and Zhang (2020) choose the number of articles referring to algorithm entities as mention units. They analyzed the academic influence of algorithm entities in the NLP field and explored the evolution of influence over time. Pan, Yan, Wang, and Hua (2015) adopted the number of sentences as the unit, and evaluated the influence of software in PLoS One by the number of sentences mentioning, using, and citing software entities. These two counting units are similar to the count X and count one method proposed by Ding et al. (2013) when evaluating the influence of references. Although the two units will bring different ranking lists, there is currently no research to analyze which one is better for method entity evaluation. On the contrary, the two influence evaluation units essentially have the same limitation: the contributions of all evaluated objects are treated equally, and as long as they appear once, they are considered valuable to the literature. This pattern cannot identify whether the method entity influence is positive or negative, nor can it distinguish more fine-grained method entities with the same counting result.

Based on the mention count, other information related to the method entities and publications can help us better understand the contexts in which such entities are involved in scientific studies. One category is the property of method entities, including the function, cost et al. Based on the function of software, Ma and Zhang (2017) obtained clusters of software entities, like statistical software, visualization software, and compared the academic influence of software with different functions. Pan et al. (2015) considered the economic properties of software entities and found that free software was more influential in the academic field than commercial software. These characteristics can lead to a more fine-grained method impact analysis, and then answer the following questions: Which type of method entity has the higher influence in a particular field? Among each sub-category (such as the classification algorithm), which entity has the higher influence? What characteristics will affect the influence of the method entity?

Another important category of such contextual information concerns the bibliographic information of published literature, such as the type of sections, publication date, discipline of the literature, and impact factors of journals. The location of the method entity (usually sections) in literature is seen as the indication of the level of influence. For example, the entities that appear in the Method section seem to be more representative and have more significant influence than entities in the Introduction section. Li (2020) studied the influence of different versions of the Diagnostic and Statistical Manual of Mental Disorders (DSM) in the method section and the non-method section in academic papers. By measuring the ratio of citations used in the method section and how this ratio shifts over time, they illustrated the extent to which a specific version of the DSM was regarded as an established research instrument. The influence of method entities in literature published in the different years is used to analyze the evolution of methods. Ding, Wang, and Zhang (2019) analyzed the influence of algorithm entities over time. They denoted that as time progressed, the influence of algorithm entities in the NLP field increased year by year, and algorithms can be classified into three types according to the trend of influence. The discipline of literature where method entities are mentioned enables the comparison of method influence between different disciplines (Duck et al., 2013; 2016). On the basis of discipline, Pan et al. (2019) compared the influence of the software entity in different disciplines and found that the software entity had a stronger influence in agriculture and biology than in social sciences and mathematics (Pan, Yan & Hua, 2016). Subsequently, they further analyzed the influence of software entities at three levels, namely literature, journals, and domains, and explored the spread of influence (Pan, Yan, Cui & Hua, 2018). These works have conducted preliminary explorations on the level, scope, and trend of the influence of method entities, reflecting the importance or trend of method entities in different contexts. But the reasons behind these phenomena have not been carefully discussed yet. The combination of these contextual information and method entity characteristics may be a good choice for exploring the reason. It can study whether the difference of method entity influence is caused by the attributes of method or is affected by the development of the overall environment.

## 5.2 Social influence evaluation of method entities

Method entities create value in both academia and industry, especially in the data-driven paradigm of knowledge production. For the influence of method entity in non-academic fields, altmetric is an excellent selection to measure it. Altmetrics are metrics and qualitative data that are complementary to traditional, citation-based metrics. It includes (but are not limited to) peer reviews on Faculty of 1000, citations on Wikipedia and in public policy documents, discussions on research blogs, mainstream media coverage, bookmarks on reference managers like Mendeley, and mentions on social networks such as Twitter ("What Are Altmetrics?", 2015). Therefore, the altmetric indicators can evaluate the influence of method entities from more comprehensive perspectives by measuring the number of mentions and downloads of these entities on social media, and the number of searches of entities in non-academic fields (such as social media platforms, search engines, and online communities). Current research on the social influence of method entities is concentrated on software entities, which is partly attributed to the fact that software is heavily used in non-academic contexts (Li & Xu, 2017).

The number of votes cast by scholars and developers is a common indicator of the social influence of method entities. The *Stack Overflow Global Developers Report* (Stack Overflow, 2020) is an annual survey on methods, web frameworks, and databases used in the community of computer science. Thousands of IT developers across the world vote for the most popular methods, web frameworks, and databases. Another Programming Community index, *TIOBE*, considers the numbers of online courses and third-party vendors, besides votes from professional programmers, and builds an index for the popularity of different programming languages based on these three measurements (TIOBE-Index, 2018). Thelwall and Kousha (2016) stated that downloads of open-source software could be used as an evaluation indicator of software influence implemented into the Depsy project (Priem & Piwowar, 2016) analyze the influence of software in the open-source community. Depsy offers both bibliometric and altmetric indicators, including the number of software downloads and the number of academic literature citations.

As a pair of complementary indicators, altmetric and bibliometric indicators can be together to deepen our understanding of the impact of method entities. For example, Zhao and Wei (2017) obtained the number of software downloads in the Depsy, citations of software in academic papers, and the dependency relationship between each software to measure the impact evaluation of open-source software. The three indicators supplement each other by focusing on different aspects of the impacts of software entities, which evaluated the domain influence, social influence, and influence among software entities, respectively. Using altmetric indicators and bibliometric indicators also reveals the relationship between the influence of method entities and literature. Yang, Huang, Wang, and Rousseau (2018) proposed a usage-based model with various indicators, including citations, mentions, and downloads to measure the importance of scientific software in the bioinformatics field. At the same time, they obtained the number of citations of each article use these software, and analyzed the correlation between the importance of software and quality of article. Their results show that scientists in the field of bioinformatics rely heavily on scientific software, and better scientific software helped produce better science.

## 5.3 The summary of method entity evaluation

Table 2 summarizes the indicators used to evaluate the impacts of method objects. The most commonly used bibliometric indicators include the numbers of citations, mentions, and the amount of use. Each indicator can be used on either the publication- or sentence level. For the altmetric indicators, the number of downloads and votes are used more frequently. Bibliometric and altmetric indicators create a complementary measurement that can evaluate the influence of method entity in different domains, and both of them are based on frequency; by default, the higher the frequency, the higher the influence or popularity.

Compared with traditional expert evaluation (Wu et al., 2018), using bibliometrics and altmetric indicators is a direct and simple way of evaluation, and it does not require high labor costs and has higher efficiency. However, as Garfield ( 1963) warned in the early years, "It is preposterous to conclude blindly that the most cited author deserves a Nobel prize." Nor can we just use frequency to measure the value of a method entity. Since each mention of a method entity has the motivation, such as introducing or using the method, and for each use of a method entity, there is a difference between direct use and improved use. Similarly, whether each mention of a method entity is affirmative or critical cannot be seen only by numbers. The combination of semantic features and frequency indicators will more accurately measure the influence of the method entity.

**Table 2. The indicators for method entities influence evaluation**

| **The types of indicator** | **Indicator** | **Counting unit** | **Reference** |
|---|---|---|---|
| **Bibliometric indicators** | The number of citations | Articles | Ding, Wang and Zhang (2019) |
| | The number of mentions | Articles | Wang and Zhang (2020), Li, Chen and Yan (2019), Li, Rollins and Yan (2017) |
| | | Sentences | Ma and Zhang (2017) |
| | The numbers of citations and mentions | Sentences | Pan, Yan, Wang and Hua (2015) |
| | | Articles & Sentences | Pan, Yan and Hua (2016) |
| | The numbers of citations and mentions and usage statistics | Articles | Pan, Yan and Hua (2019) |
| **Altmetric indicators** | The number of votes | —— | Stack Overflow (2020), TIOBE-Index (2018) |
| | The number of downloads | —— | Thelwall and Kousha (2016), Priem and Piwowar (2016) |
| **Altmetric indicators &bibliometric indicators** | The number of downloads & the number of citations & the number of mentions | Articles | Zhao and Wei (2017), Yang, Huang, Wang and Rousseau (2018) |

At the same time, according to table 2, it can be seen that prosperities and context information are added to analyze the influence of different methods in different fields, which means that the author is no longer just satisfied with getting a simple ranking list of method entities. They did find exciting conclusions. For example, different disciplines have different requirements for specific method entities, and there is a correlation between suitable methods and good research results. Nevertheless, these phenomena obtained through influence analysis currently lack in-depth analysis. Maybe we should shift these quantitative analysis results to qualitative analysis and understand the reason behind the phenomenon. In this way, the evaluation of method entities can bring theoretical contributions to the use of methods and the development of the field.

## 6. Application of method entities from academic literature

Method entities collected from academic literature have strong technical and social relevance that can be further utilized in user applications. Applications include the following categories, the corpus of method entity, and the method entity retrieval platform. These applications provide a display platform for a large number of method entities. They support users to directly download well-labeled method entities for subsequent automatic extraction and other research. Users can also retrieve methods based on method names or task names in these applications to learn and select methods. This section will introduce these two types of application platforms based on method entities.

### 6.1 The method entity corpus and database

A large number of method entities obtained from academic literature are important resources for the method entity corpus and database. These corpora contain at least a series of independent method entities, and the richer corpus will also provide the sentence or article content where the method entities are located. Handschuh, QasemiZadeh and Schumann constructed the technical term corpora *ACL RD-TEC* (Handschuh & QasemiZadeh, 2014) and *ACL RD-TEC 2.0* (QasemiZadeh & Schumann, 2016), providing a vital method entity dataset based on ACL conference articles. In the first version, they marked documents in the ACL-ARC (Bird et al., 2008) and generated candidate terms using n-gram techniques. From these candidate terms, the authors manually identified technical entities used in the field of computational linguistics, including methods, algorithms, and solutions. In version 2.0, taggers annotated terms in abstracts of ACL conference proceedings, based on which the corpus was built. All terms were classified into seven categories: *method, tool, language resource (LR), LR product, model, evaluation, and others*. The two corpora provide the data for entity recognition and term evaluation tasks.

Some algorithmic programming competitions also provide corpora of method entity, including the name of methods and the sentence with labeled method entity. *SemEval*[2] is a series of international natural language processing (NLP) research workshops whose mission is to advance the current state of the art in semantic analysis and help create high-quality annotated datasets. In *SemEval 2017*, task 10 required participants to automatically extract entities such as scientific processes and materials from five hundred scientific publications from the domains of Computer Science, Material Sciences, and Physics (Augenstein et al., 2017). Therefore, the organizers provided 500 paragraphs from journal papers evenly distributed among the three domains, and keywords about the task, process, and material in these paragraphs were labeled with exceptional marks. Based on SemEval 2017, Luan, He, Ostendorf, and Hajishirzi (2018) created the dataset SCIERC, which includes annotations of scientific terms, relation categories, and co-reference links from abstracts of 500 scientific publications. These publications were taken from 12 AI conference/workshop proceedings from the Semantic Scholar Corpus. Scientific terms in the abstract of the 500 articles were labeled and classified into Task, Method, Evaluation Metric, Material, Other Scientific Terms, and Generic terms. Unlike datasets that contain multiple classes of method entities, Du et al. (2021) published a dataset that only targets software entities. They manually annotated 4,971 full-text research articles published during the period 2000–2010, of which 2,521 were from the field of biomedicine and 2,450 were from the field of economics. A total of 4,093 software entities were identified from 1,228 articles, and 2,541 software were mentioned with their details including publisher, version, and URL.

[2] https://semeval.github.io/

Information about these corpora and databases is displayed in Table 3. The currently available method entity corpora are mainly concentrated in computer science, and the amount of data in different corpora varies from hundreds to tens of thousands. Structured full-text content and labeled entities provide training data for automatic method entity extraction, but we are unsure whether these corpora are of high quality. At the same time, corpora are constructed based on publications from different disciplines. Whether the method entity corpus of a specific field can be used for entity extraction in other fields still needs further discussion.

**Table 3. Information about different method entities corpus**

| Corpus | Year | Field | Scale | URL |
|---|---|---|---|---|
| ACL RD-TEC | 2014 | Computer science | 10,922 papers with 13,832 technical terms | https://github.com/languagerecipes/the-acl-rd-tec |
| ACL RD-TEC 2.0 | 2016 | Computer science | Abstract of 300 papers with 6,818 terms | https://lindat.mff.cuni.cz/services/kontext/run.cgi/first_form?corpname=aclrd20_en_a. |
| SemEval2017 | 2017 | Computer science, Materials, Physics | 500 papers and keywords | https://scienceie.github.io/resources.html |
| SCIERC | 2018 | Computer science | 500 scientific abstracts with 8,089 methods, tasks, metric entities | http://nlp.cs.washington.edu/sciIE/ |
| Softcite | 2021 | Biomedicine, economics | 1,228 papers with 4,093 software entities | https://zenodo.org/record/4445202#.YhQvTc_ityw |

## 6.2 Method entities retrieval platforms

The method entity retrieval platforms are built on method entities and related resources and information, such as tasks, data sets, and academic documents. According to the method entity word searched by the user, these platforms provide users with the definition of the method, the results of the method in different tasks, the papers, monographs about the method, and others. Based on the search results provided by the platform, we divide existing platforms into three categories and introduce some representative examples in this section.

The first type of platform provides users with learning resources related to method entities, and *SAGE Research Methods*[3] is one of the representatives. SAGE Research Methods is a platform that focuses on research methods in various disciplines. When the user searches the name of a research method, the platform will provide resources such as books, videos related to the method entity, data sets analyzed by the method, et al. All results can be further filtered according to type, discipline, and publication time.

The second type of platform is constructed based on academic literature, providing users with the original sentences where method entities are cited or mentioned. As exemplified by the Method Library of China National Knowledge Infrastructure (CNKI), they provide a list of commonly used disciplinary methods in some example disciplines, which the user can click directly to obtain the content of the method. In addition, users can also retrieve the method entities they want to know. No matter which pattern, the platform will feedback a series of sentences containing method entities (searched or clicked by users) extracted from academic literature. These sentences are divided into the creator, creation time, definition, development process, principle, application, sub-category, feature, step, and function of the searched method. Users can view these sentences by category and read the paper where the sentence is located through the attached link. Another example is the *IBM Science Summarizer*[4]. Unlike the CNKI method library, IBM does not return original sentences but a summary of the content of the section where the method entity is located. The developers used dictionary matching and machine learning technology to identify tasks, data sets, and evaluation metrics in 270,000 papers from arXiv.org (“Computer Science” subset) and ACL anthology. The platform summarized sections in different papers that mentioned the method entity being retrieved, allowing users to focus on the relevant sections for the task at hand (Erera et al., 2019).

The third type of platform display the ranks extracted method entities based on their performance evaluated by metrics (accuracy, recall, F-value) in different task. The *"SOTA" project at Heart of the Machine* is a notable example. Developers classify the method entities by the task that can be solved by the method. When users search for a task, they can get a list of method entities, each of which belongs to an existing paper. At the same time, the dataset used by each method entity and the result achieved by the method entity that evaluated by the accuracy, recall, F-value are also shown with the method. Additionally, users can know the state-of-the-art (SOTA) model and the most commonly used model in each task. Another good example is *Papers with Code*[5], with the similar principles and functions of "SOTA" project, Paper with code allow users to retrieve method entities according to the task, and then give users the state-of-the-art model, the data set used, and the F-1 value achieved by the SOTA model. In addition, they provide the direct search function of the method; that is, the user can not only retrieve the task to obtain the method but also retrieve the method to obtain the related information. The platform will return the definition of the method and the paper containing the method in the title. Another highlight is that Paper with Code establishes a link between academic papers and academic communities. If authors of a paper share their code on Github[6], paper with code will also include the code of method entities in the retrieval result.

[3] http://methods.sagepub.com/
[4] https://ibm.biz/sciencesum
[5] https://paperswithcode.com/

Table 4 summarizes the platforms mentioned above. In general, these platforms have a similar goal and aim to help scholars better understand, learn methods and choose appropriate methods for their own tasks. However, feedback on retrieval in these platforms is different: SAGE provides the most prosperous search results, more like a knowledge database about methods. CNKI and IBM seem to be academic intermediaries, and they are responsible for delivering original or rough-processed content, helping users understand the context of method entities in academic literature. Heart of the Machine and Paper with Code have conducted in-depth processing of the content of the academic literature related to the method entity. They have to extract other entities related to the method entity in the academic literature, namely, the task, data set, result, and rank the method entities by the experiment result in each task. The first two types of platforms are method-driven platforms. Users must know what methods they want to understand and learn before searching. The third type of platform can be operated in a problem-driven mode, which means users do not need to know what method they want but find a method through the problem or task they need to solve. Therefore, combining these three types of platforms may be a better choice. First, find the most classic and latest method entities based on the task, then find out the content of a large number of method-related academic papers on the second type of platform, and finally obtain learning videos or monographs on the first type of platform to promote the understanding of the method.

**Table 4. Information about different method entity platforms**

| Platform | Year | Developer | Function | Field | Scale |
|---|---|---|---|---|---|
| SAGE Research Methods | 2014 | SAGE press | According to the search terms, SAGE supports users searching sociological research methods, and then feedback books, reference books, manuals, journal literature, cases, teaching videos, and university survey data. | Social science | More than 1,000 monographs and papers, 770 method terms |
| Papers With Code | 2019 | Facebook AI | Users can access relevant papers and corresponding codes via the title keywords and arrange "hot research" by popularity and GitHub stars. | Machine learning field | More than 60,000 academic papers |
| IBM Science Summarizer | 2019 | IBM | Based on the user's retrieval needs, the system feedbacks the chapter level abstracts, full-text level abstract, task, data set, and evaluation index of a single paper to help users quickly understand the article content. | Computer science | 27,000 papers, 872 tasks, 345 data sets, and 62 metrics entities |
| The CNKI method library | 2019 | CNKI | Users can query from the perspectives of "method". The system will provide the sentences in different types that contain the searched entity. | Multi-field | Academic papers in CNKI |

[6] https://github.com/

| | | | | | |
|---|---|---|---|---|---|
| "SOTA" project | 2019 | Heart of the machine | Users can directly search for tasks in the machine learning domain to obtain the task definition, SOTA model name, the most commonly used model, and the most popular data set. | Machine learning field | More than 100 concepts, technical methods, evaluation indicators, and results of hundreds of machine learning tasks |

# 7. Discussion and conclusions

The large-scale extraction, evaluation, and application of method entities from academic literature can help scholars to understand method entities in specific fields comprehensively. Based on these studies, scholars can carry out a series of method entity analyses, including proposing and evaluating new methods based on the existing methods, analyzing the relationship between methods, constructing the framework of methods in a field, predicting the influence of methods, exploring the evolution of a discipline by method, establishing interdisciplinary collaboration through methods, recommending methods for beginners and other researches. In the current work related to method entities, there are still problems that have not yet been resolved or need to be optimized. This section will discuss these limitations and propose open questions that can be studied in the future.

## 7.1 Limitations to research on method entities in academic literature

Given the current research on method entities in academic literature, the following limitations still need to be explored.

**(1) The source of method entities is limited**

The first source limitation means the object of large-scale method entity research is mainly methods in the field of natural sciences. As Khan et al. ( 2017) stated, the most significant limitation of existing big scholarly data platforms is that they are subject-specific: most of them only cater to the computer science community. The same limitation applies to studies on method entities, which is that scholars pay more attention to the methods that appear in a limited array of research fields like computer science, biomedicine, and information science. We rarely see the work of extracting a large number of social science methods or humanities and artistic methods. What are the characteristics of the method entities extracted from the literature of different disciplines? Which methods are shared among multiple disciplines, and which methods are unique to the discipline? No corresponding answers have been found to these questions.

The second source limitation refers to the fact that most of the current method entities come from academic papers. Other types of academic literature, such as monographs, are rarely used to extract method entities. However, the academic value of these documents in method entity analysis cannot be ignored, for example, in the field of humanities and social sciences. Academic monographs are considered to be more influential than academic papers. It is possible that we can find more classic method entities with greater influence from the monographs.

**(2) The performance of method entity extraction needs to be improved**

With the development of science and technology, the extraction technology of method entities has experienced a transformation from manual rules to statistical machine learning. In order to understand the current achievements of scholars in method entity extraction, we compared the best results of current genera NER work and that of method entity extraction work based on the F-1 value. In the *paper with code*, we found the current best NER result, implemented by the ACE + document-context model on CoNLL 2003, and it reached an F-1 value of 94.6%. At the same time, we checked the best results in the current method entity extraction work. For general methods, the best results are 78.26% (in sentences) and 68.50% (in paragraphs), respectively. For a specific method object, the best results are 88.4% in software entity extraction (in the full text of academic papers).

We speculate that the above difference in results is related to the type of object to be extracted. It has been proved that biomedicine NER showed less performance than general NERC systems due to the complex structure of biomedical entities (Shen et al., 2003). Therefore, textual genres or domain factors may also cause the method entity extraction to be inferior to the general NER result. Because in the process of literature review, we found that the definition of methods and method entities is challenging to clarify, and the definitions and scopes given in different related works are not the same. Compared with CoNLL2003, which only requires the extraction of person, organizations, and locations from the news (Tjong Kim Sang & De Meulder, 2003), it is a big challenge for the machine to automatically extract the method entity from the academic literature with such a complex definition.

In addition, most of the current method entity extraction stops until the method entity is acquired, and further clean-up work is not carried out. Only a tiny amount of work has normalized entities through manual review (Wang & Zhang, 2020) or rules (Luan, 2018) where different words representing the same method were summarized into a set and constructed a method dictionary. However, another critical issue has not been resolved, namely, the semantic disambiguation of method entities. In academic papers, method entities are often expressed as abbreviations, and the same abbreviation may represent different methods. For example, "BP" can mean both the "Belief propagation" algorithm and "Back propagation" algorithm. If the full name of abbreviations in each sentence or document can be disambiguated, it can improve the performance of method entity recognition and subsequent evaluation.

**(3) The singular evaluation of method entities**

Comprehensive and objective evaluation of scientific entities is not an easy study, and existing method entity evaluation also faces this problem of singular evaluation. Although the authors try to combine different metrics to evaluate the influence of method entities, metrics alone have been unable to achieve the task of predicting scientific impact and assessing research quality (Sahel, 2011). Because the nature of metrics used for method evaluation is the ranking based on frequency, in this way, only the classical method entities stand out because they appear in the article with high frequency over time. Classical methods are critical for developing a field, but emerging methods in a field are also worthy of attention. These rising stars may not have accumulated long-term and far-reaching influence like the classic methods, but they may have a strong influence in a short period and lead the direction of future research. However, the existing method impact assessment work is difficult to find these meaningful new methods that are likely to become the classic methods in the future.

In addition, existing evaluation work has tried to combine frequency with method characteristics and contextual information to discover differences and changes in influence between method entities.

Nevertheless, it may be more necessary to combine frequency with semantic information. The current evaluation treats every mention of the method equally, but semantics can help scholars judge the difference in each mention and the difference in influence between method entities with the same influence. Of course, there is no uniform standard for classifying or weighting mentions based on semantics. How to prove the rationality of the classification system or weighting will become a new problem. Perhaps this is why few studies have tried to evaluate the influence of method entities based on semantics.

## 7.2 Future directions

To sum up, research on method entities in academic literature is a rapidly developing field of research. We have seen valuable work on the extraction, evaluation, and application of method entities. At the same time, we also believe that there are many issues about the method entity worthy of further study. Therefore, in this section, we discuss the future directions of method entity-related research

**(1) Carrying out large-scale method entity extraction with the limited corpus**

Compared with the data set for general NER, the public data set of method entities has a small number and a single type. However, the current popular deep learning models often require large-scale training data, which makes less attention on method entity extraction, and the performance of the extraction model is not as good as that in the general NER. This problem can be solved from two perspectives.

First of all, try to construct a richer method entity corpus, such as annotating method entities in the text through crowdsourcing, but it will inevitably invest much workforce. Therefore, it is also possible to use the mode of combining machine process and manual review. Using the existing corpus to train the machine learning model to extract more candidate entities, the reviewer can manually find the method entities. After that, matching the new method entities with new text can be used to generate a new training corpus. In addition, researchers can also try to use machine translation to generate corpus in other languages based on existing training corpus in a single language. Secondly, if scholars do not want to invest too much energy in constructing the corpus, they can also reduce the model's dependence on large-scale training corpus. Both few-shot learning and transfer learning approaches are worth trying. To reduce manual input, selecting a suitable model and using a training method that is less dependent on data to obtain better extraction results is one of the future directions for method entity extraction.

**(2) Conducting fine-grained evaluation on method entity**

In the future, the evaluation of method entities in academic publications could consider multiple perspectives of the impact of method entities, including the influence, performance, cost, et al.

Influence can be subdivided into academic influence, social influence, and technological influence. Scholars can try to use more types of frequency indicators. For academic influence, they can create a new indicator to find rising stars, and use the number of disciplines to measure the universality of the method. For social influence, in addition to the number of downloads, the number of mentions in social media, such as the number of tweets, and the number of reposts of method-related papers, can also be used as new altmetric indicators. For technological influence, scholars can consider mining and evaluating the method entities mentioned in the patent. At present, the existing literature analyzed the citation of academic papers in patents to measure the technical influence of academic papers (Veugelers & Wang, 2019). Similarly, the number of mentions and citations of a method entity in patents can also be used to indicate the technological influence. For the above three types of influence, we can also analyze whether there is a correlation between the three types of influence for the same method entity.

Performance evaluation refers to measuring the result of method entities in a specific task, such as the precision, recall and F-value achieved by different algorithms in classification tasks. The traditional approach requires specific experiments to observe the results achieved by different method entities in the task (Settouti et al., 2016), which is time-consuming. By the way, many authors will share their methods and results in their papers, so we can try machine learning methods to automatically identify the experimental results of each method entity and then classify these methods according to the task they solve. In this way, the performance of different method entities in the same task can be evaluated with less workforce.

Cost evaluation refers to the measurement of various costs involved in the use of the method. Taking deep learning methods as an example, we can measure the scale of training corpus they need in research, the equipment investment, and even energy consumption. We should pay attention to the cost of the method because the energy consumption of training a deep learning model may be as high as 626,155 pounds of carbon dioxide emissions (Strubell, Ganesh & McCallum, 2019). Performance evaluation can let people understand the output and effect of the method, and what resources these outputs consume. We always hope to get more methods with little investment and high performance.

**(3) Exploring the relationship between method entities**

There are different relationships between method entities, and these relationships can be used to construct method frameworks or method knowledge graphs in different disciplines, fields, and tasks. Existing work has made use of the co-occurrence relationship between methods, explored the evolution path of methods (Duck, Kovacevic, Robertson, Stevens & Nenadic, 2015; Zha et al., 2019), and constructed a knowledge graph between methods, tasks, and results (Dessì et al., 2020; Luan et al., 2018). However, there is still much related work that can be explored in the future.

The first type of work is to explore more fine-grained relationships between method entities. We can use contextual semantics to understand whether the methods appearing in the same task are compared as competitors, combined as partners, or new methods are generated based on old methods. Subsequently, we can get a method framework or graph that includes the upper-lower and peer relationships. The method entities that have relations in the graph will be connected, while the methods that are not connected are likely to generate new method combinations to solve research problems in the future, which is worthy of our attention. In addition, we can generate method entities in a specific field in different periods to understand the changes in the overall method structure in the field, speculate the reasons, and interpret the development of the field or discipline from the perspective of methods.

The second type of work is to explore the relationship between method entities and other entities. Considering the method entity and discipline (task) entity simultaneously, we can obtain the diffusion path of the method entity in different disciplines (task). In addition to common tasks and results, we can also use methods and the person who are the users of a method to build a knowledge graph together. In this graph, we can clearly see who used what methods to solve what problems and get what results. The method entity can be expressed as the scholar's skills that can be used to construct a richer scholar portrait together with the basic information of scholars. The method can also be the research interests of different scholars, and the research interest can be combined with the relationship of methods to recommend collaborators with the same or complementary interests for different scholars.

### (4) Developing platforms for method retrieval and recommendation

When we try to find method entities based on tasks from existing method retrieval platforms, we can only get the method entity that produces the final result on the public data set. However, solving a task is a complex process: not all tasks deal with the same data, and users themselves have different requirements for methods. In solving a task, we may need to disassemble the task into multiple steps, and each step corresponds to a different subtask and requires a different method. These different methods constitute a complete method system to solve the task in order. Combining the method with different data sets and results can create a complete task-solving framework. Future retrieval platforms should first try to mine all method entities involved in different tasks, and then analyze the relationship between method entities and other entities. Next, developers need to find out the logical relationship between all the entities applied in the process, from raising the problem to solving the problem and transforming the task-solving process into a complete structured framework. After that, results of multi-dimensional method evaluation can be utilized to describe the characteristics and performance of method entities appearing in the framework. Users can obtain the landscape of solving various tasks, including the steps to solve the task, the methods used in each step, and the final results. We hope that the systematic results found on the platform can provide valuable recommendations for authors to find the most suitable.

## Acknowledgment

This study is supported by the National Natural Science Foundation of China (Grant No. 72074113).

## **Appendix:** The database of academic literature used for method entity extraction

| Database | Coverage | File format | Link |
|---|---|---|---|
| Web of science (WOS) | WOS Core Collection contains over 21,100 peer-reviewed, high-quality scholarly journals published worldwide in over 250 sciences, social sciences, and arts & humanities disciplines; provides 1.9 billion cited references from over 171 million records. | Bibliographic content of academic literature in TXT format | https://www.webofscience.com/wos/woscc/basic-search |
| ACL Anthology | The ACL Anthology hosts 66,259 papers from conferences and journals in the domain of computational linguistics and natural language processing. | (1) Full-text content of academic papers in PDF format<br>(2) Title, author, and abstract of academic papers in BibTex format | https://www.aclweb.org/anthology/ |
| ACL Anthology Reference Corpus (ACL ARC) | The ACL ARC contains 16,718 papers from 10 conferences in the domain of computational linguistics and natural language processing. | Full-text content of academic literature in XML format | https://acl-arc.comp.nus.edu.sg/ |
| Pubmed Central | PMC contains 7.1 million articles from 10,775 journals, spanning several centuries of biomedical and life science research (the late 1700s to present). | Full-text content of academic literature in XML format | https://www.ncbi.nlm.nih.gov/pmc/ |
| PLoS | PLoS contains 12 open access journals and more than 200,000 articles | Full-text content of academic literature in both PDF and XML format | https://plos.org/ |